\documentclass[final]{siamart220329}

\usepackage[english]{babel}
\usepackage[T1]{fontenc}
\usepackage[utf8]{inputenc}

\usepackage{booktabs}
\usepackage{graphicx}

\usepackage{xcolor}
\usepackage{csquotes}
\usepackage{hyperref}

\hypersetup{colorlinks=true}
\hypersetup{linkcolor=black}
\hypersetup{citecolor=black}

\usepackage{amsmath}
\usepackage{amssymb}
\usepackage{physics}
\usepackage{nicefrac}

\renewcommand{\vec}[1]{\boldsymbol{#1}}
\newcommand{\mat}[1]{\boldsymbol{#1}}

\DeclareMathOperator{\diag}{diag}

\newcommand{\meq}{\mathrm{eq}}

\newcommand{\abc}{\alpha\beta\gamma}

\newcommand{\pystencils}{\emph{pystencils}}
\newcommand{\lbmpy}{\emph{lbmpy}}
\newcommand{\waLBerla}{\emph{waLBerla}}
\newcommand{\SymPy}{\emph{SymPy}}

\usepackage{xcolor}

\definecolor{lssblue}{HTML}{1F3B66}
\definecolor{lssred}{HTML}{C24D30}
\definecolor{lssgreen}{HTML}{1F6622}

\usepackage{listings}

\lstdefinestyle{pysnippet}{
  numbers=left,
  captionpos=b,
  breaklines=true,
  xleftmargin=0pt,
  frame=tb,
  language=Python,
  showstringspaces=false,
  basicstyle=\scriptsize\ttfamily,
  keywordstyle=\bfseries\color{green!40!black},
  commentstyle=\itshape\color{purple!40!black},
  identifierstyle=\color{black},
  stringstyle=\color{orange},
  mathescape=true,
}

\crefname{lstlisting}{listing}{listings}
\Crefname{lstlisting}{Listing}{Listings}

\usepackage[textsize=footnotesize,textwidth=4.0cm]{todonotes}
\newcommand{\todURi}[1][]{\todo[color=green!20, author=Uli, size=\footnotesize, inline]}
\newcommand{\todUR}[1][]{\todo[color=green!20, size=\footnotesize, linecolor=darkgray, author=Uli]}
\newcommand{\todFH}[1][]{\todo[size=\footnotesize, linecolor=darkgray, author=Frederik, inline]}
\usepackage{booktabs}
\usepackage{multirow}

\usepackage{siunitx}

\usepackage{acro}

\DeclareAcronym{PDE}{
    short=PDE,
    long={partial differential equation}
}

\DeclareAcronym{NSE}{
    short=NSE,
    long={Navier-Stokes equations}
}

\DeclareAcronym{SWE}{
    short=SWE,
    long={shallow water equations}
}

\DeclareAcronym{LBM}{
    short=LBM,
    long={lattice Boltzmann method}
}

\DeclareAcronym{LB}{
    short=LB,
    long={lattice Boltzmann}
}

\DeclareAcronym{SRT}{
    short=SRT,
    long={single relaxation-time}
}

\DeclareAcronym{BGK}{
    short=BGK,
    long={Bhatnagar-Gross-Krook}
}

\DeclareAcronym{TRT}{
    short=TRT,
    long={two relaxation-time}
}

\DeclareAcronym{MRT}{
    short=MRT,
    long={multiple relaxation-time}
}

\DeclareAcronym{API}{
    short=API,
    long={application programming interface}
}

\DeclareAcronym{CQE}{
    short=CQE,
    long={conserved quantity equations}
}

\DeclareAcronym{CQC}{
    short=CQC,
    long={conserved quantity computation}
}

\DeclareAcronym{CSE}{
    short=CSE,
    long={common subexpression elimination}
}

\DeclareAcronym{SSA}{
    short=SSA,
    long={static single assignment}
}

\DeclareAcronym{GPU}{
    short=GPU,
    long={graphics processing unit}
}

\title{%
    Advanced Automatic Code Generation for Multiple Relaxation-Time Lattice Boltzmann Methods
}

\author{%
    Frederik Hennig%
    \thanks{Chair for System Simulation, Friedrich-Alexander-Universität Erlangen-Nürnberg,
            Erlangen, 91058 Germany
            (\email{frederik.hennig@fau.de}),}
    \and Markus Holzer\footnotemark[1]{}
    \thanks{CERFACS, 42 Avenue Gaspard Coriolis, 31100 Toulouse, France (\email{holzer@cerfacs.fr}).}
    \and Ulrich Rüde\footnotemark[1] \footnotemark[2]
}

\headers{Advanced Lattice Boltzmann Code Generation}{Frederik Hennig, Markus Holzer, and Ulrich Rüde}

\begin{document}
    \maketitle

    \begin{abstract}
The scientific code generation package \lbmpy{} supports
the automated design and the efficient implementation of \acfp{LBM} through metaprogramming.
It is based on a new, concise calculus for describing \acl{MRT} \acp{LBM},
including techniques that enable the numerically advantageous subtraction of the constant background component
from the populations.
These techniques are generalized to a wide range of collision spaces and equilibrium distributions.
The article contains an overview of \lbmpy{}'s front-end and its code generation pipeline,
which implements the new \ac{LBM} calculus by means of symbolic formula manipulation tools and object-oriented programming.
The generated codes have only a minimal number of arithmetic operations.
Their automatic derivation rests on two novel Chimera transforms
that have been specifically developed for efficiently computing raw and central moments.
Information contained in the symbolic representation of the methods is further exploited in a customized 
sequence of algebraic simplifications, further reducing computational cost.
When combined, these algebraic transformations lead to concise and compact numerical kernels.
Specifically, with these optimizations,
the advanced central moment- and cumulant-based methods can be realized with only little additional cost
as when compared with the simple \acs{BGK} method.
The effectiveness and flexibility of the new \lbmpy{} code generation system is demonstrated in simulating 
Taylor-Green vortex decay and the automatic derivation of an \ac{LBM} algorithm to solve the \acl{SWE}.

    \end{abstract}


    \section{Introduction}
\label{section:Introduction}

Modern scientific computing is characterized by growing complexity on several fronts.
On the one side, the mathematical models become more advanced and powerful so that
the corresponding numercal methods and algorithms grow inceasingly intricate and involved.
At the same time, contemporary computer architectures become more difficult to program.
This is especially true in high end parallel computing that is characterized by rapidly evolving new hardware architectures,
such as general-purpose \acp{GPU}.

Thus the central step of scientific computing, i.e., the step from a model to executable software, 
faces increasing challenges.
The developers of algorithms and scientific software are squeezed in from two sides: 
Increasingly  complex simulation models must be realized in software for increasingly complex advanced
supercomputer systems. 
We note here also that legacy software often lacks performance portability.
For example it may present an almost prohibitive effort to port a legacy flow solver to a modern multi-GPU system.
It may be speculative, but likely enormous computational resources are underused world wide, since outdated software
is underperforming on the given computer systems on which it runs.

At the same time, we observe that modern mathematical methods have evolved to a state where their derivation and analysis
depends increasingly on researchers making use of symbolic formula manipulation packages. 
For the present article, we will consider the class of \acfp{LBM}, whose most advanced variants could hardly be
derived without using symbolic mathematical software.
Given this situation, it is a natural step to integrate the algebraic manipulations routinely into the workflow of
designing and implementing these methods. 
A first step of this is presented for the \lbmpy{} package by Bauer et al.\ in \cite{Bauer2021lbmpy}.

We emphasize that the effect is twofold: In this approach, the derivation of the methods is consolidated by giving method developers
a powerful tool to describe, analyze, and experiment with a range of method variants.
But at the same time, the automated symbolic derivations can be used beyond just deriving the discrete approximations.
Conventionally, this would then only be another set of mathematical formulas that still have to be
transformed into software manually.
Here, instead, the symbolic manipulation will be taken a step further to produce operational code directly.
Through this automatic code generation approach, numerical kernels can be produced
targeting different architectures (such as CPUs or \acp{GPU}),
using different data structures, and also different memory layouts.

Furthermore, symbolic manipulation offers the possibility to optimize the codes in various ways,
both mathematically by simplifying expressions,
but also with hardware aware transformations, e.g., with the goal of providing an automatic vectorization.
A specific advantage here is that the optimizations can go much beyond traditional optimizing or vectorizing compilers.
The domain-specific code generator may leverage information contained in the numerical method's symbolic form
to apply highly specific simplifying transformations.
This enables automatic optimizations that would otherwise be accessible only to an expert programmer who is at the same time
a highly educated model developer ---a situation which in our experience is  rarely found  in scientific computing practice. 
In \lbmpy{} this rare synthesis of expertises is encapsulated within the automatic code generation package itself.

The \ac{LBM} is a mesoscopic method for the simulation of transport phenomena,
originally developed as an extension of lattice gas automata \cite{Chen1992}.
More modernly it is derived by the discretization of the Boltzmann equation \cite{Krueger2017}.
It discretizes the continuum of a fluid on a cartesian lattice, modelling its state
by the local distribution of particle velocities at the lattice sites.
The first application of the  \ac{LBM} was as a solver of the \ac{NSE},
but the \ac{LBM} has also been applied to a variety of problems.
These include the advection-diffusion equation \cite{Krueger2017},
thermal flows \cite{Duenweg2007},
the \acl{SWE} \cite{Zhou2002,Rosis2017ShallowWater,Venturi2020CascadedSWE},
as well as multiphase and multicomponent flows \cite{Krueger2017,Geier2015PhaseField,Holzer2021,Gruszczynski2020,Sitompul2019}.

At the core of the \ac{LBM} algorithm lies the \emph{collision} step,
where the local particle distributions are relaxed towards a given equilibrium distribution.
The equilibrium state depends on the macroscopic physics that the method is meant to simulate, and thus
differs significantly between applications.
The classical variant of the relaxation process is the discrete \ac{BGK} collision operator \cite{Bhatnagar1954,Chen1992,Krueger2017},
which employs a single relaxation rate.
It has been generalized to \acf{MRT} collision operators, relaxing different statistical moments of the distributions
with individual rates, thus improving the representation of simulated physics, as well as numerical stability
\cite{dHumieres2002MRT,Ginzburg2008,Geier2006,Geier2015,Krueger2017}.
Originally employing raw moments as a collision space,
\ac{MRT} methods were later extended to central moment \cite{Geier2006,Premnath2011CM3D,Rosis2017}
and cumulant \cite{Geier2015} space.

As a complex numerical method, the \ac{LBM} is subject to floating-point roundoff error,
which may pollute solutions when the floating point precision is insufficient and not properly controlled.
Then the accuracy of the simulation may deteriorate \cite{Skordos1993,Geier2020,Lehmann2022}.
Early in the development of \acp{LBM} it has been realized that, in hydrodynamic simulations, populations
deviate only little from a background distribution typically given by the lattice weights \cite{Skordos1993}.
Since this background distribution is invariant under collision,
it may be subtracted from the population vector, thus significantly increasing the number of digits available
for an accurate floating point presentation of the populations.
We use the term the \emph{zero-centered} storage format for this algebraic transformation.
It has been found to improve the \ac{LBM}'s numerical accuracy substantially \cite{Geier2015,Geier2020,Lehmann2022}.

Implementing all  relevant variants of collision operators and application use cases in a single, generic software
framework poses a significant challenge.
The challenge becomes even greater when different hardware architectures require different optimizations.
The creation of the modular and object-oriented software design
of frameworks such as Palabos \cite{Latt2021Palabos},
OpenLB \cite{Krause2021OpenLB}, TCLB \cite{TCLB},
and waLBerla \cite{Bauer2021waLBerla,Schornbaum2016,Schornbaum2018}
is a key step to solve this problem.
However, as elsewhere, the generality of a software often inhibits specific performance optimizations.
An automatic code generation approach, as described above, constitutes a potentially more flexible alternative to an extensive but rigid hand-coded framework. 
This idea has been successfully employed in the context of classical finite element discretizations, e.g. in \cite{Firedrake, Dolfin},
or for general stencil codes \cite{Physis, SDSLc, Modesto, Exastencils}.
Metaprogramming techniques can also be found in some of the aforementioned \ac{LBM} frameworks \cite{TCLB, Krause2021OpenLB}.
The original version of \lbmpy{} adopts this paradigm in the form of a symbolic domain-specific language based on
the raw moment \ac{MRT} formalism.
Its key functionality is the automatic generation of optimized implementations of \ac{SRT} \cite{Chen1992},
\ac{TRT} \cite{Ginzburg2008} and raw-moment \ac{MRT} methods from their symbolic description.

The aim of this work is the extension of \lbmpy{} to go beyond the existing functionality and support
both central moment and cumulant collision spaces,
to integrate the zero-centered storage format,
and to provide the transformations needed to generate \ac{LBM} codes for various applications beyond the \acl{NSE}.
During the development process, we found the original structure of \lbmpy{} to pose severe restrictions.
Therefore it was necessary to re-design the original architecture of \lbmpy{}.
The outcome of this process will be presented in this article.
It has three essential components:

First, we introduce a consolidated theoretical framework for modelling \ac{MRT} methods.
This defines a generalized formalism to specify the collision spaces and the equilibrium.
Additionally, as a novel feature, this framework integrates the zero-centered representation of the distribution functions.
In particular, we develop and present a generalization of zero-centered storage to collision spaces,
and formalize its application to the equilibrium distribution.

Second, we present the front-end of \lbmpy{} as an \ac{API} in the Python programming language
for modelling \acp{LBM} on an abstract level.
This constitutes a domain-specific language that closely reflects the theoretical framework of \acp{LBM}.
It is implemented by means of the computer algebra package \SymPy{} \cite{sympy}.

Third, we describe the modularized automatic procedure for symbolically deriving and optimizing the collision rule.
This includes the presentation of two novel Chimera transforms between populations and the raw and central moment spaces,
designed to minimize the number of arithmetic operation for the mappings between these spaces.
We furthermore elaborate on an extensive collection of \ac{LBM}-specific algebraic simplifications.
Due to these improvements, our new code generator can produce efficient implementations of raw moment,
central moment, and cumulant \acp{LBM}.
As a surprising key result we find that the latter of these advanced \acp{LBM} require only little arithmetic overhead
as compared to simple \ac{SRT} and raw moment \ac{MRT} methods. 
Thus, in many cases the advanced methods are expected to execute almost equally fast as the simpler ones.

In summary, our article presents a versatile and flexible framework for
the modelling and prototyping of complex \ac{LBM} methods,
combined with the techniques for the automatic generation of highly optimized computational kernels for these methods.
The software framework itself is open-source and freely available under the GNU AGPLv3 license.

The remainder of this paper is structured as follows.
In \cref{section:TheoreticalBackground}, we introduce our theoretical framework and present our
generalization of zero-centered storage.
\Cref{section:MRTModelling} presents the modelling front-end of \lbmpy{}.
\Cref{section:DerivationAndCodegen} elaborates on the automatic derivation procedure
for collision rules, including our novel raw and central moment Chimera transforms.
The same section introduces the extensive simplification procedure applied by \lbmpy{}
and shows its effectiveness for optimizing collision kernels.
Finally, we demonstrate the versatility of the revised framework for modelling advanced \acp{LBM},
its applicability to their rapid implementation and testing,
and the numerical advantages of zero-centering in \ac{MRT} methods
in \cref{section:Applications}.
\Cref{section:Conclusion} concludes the paper.

    \section{Theoretical Background}
\label{section:TheoreticalBackground}

This section presents the theoretical framework that serves as the basis of our code generation system.
To this end, we first introduce our formalized approach to modelling populations and the zero-centered storage format.
The multiple relaxation-time \ac{LBM} collision process is introduced and generalized to arbitrary collision spaces.
In particular, we discuss raw moment, central moment and cumulant spaces, as well as the reflection
of zero-centered storage therein.
Finally, we introduce a novel abstract interpretation of the equilibrium distributions.

\subsection{Discrete Lattice Structure and Storage Format}
\label{subsection:LatticeStructureAndStorage}

The \acl{LBM} acts on a $d$-dimensional cartesian lattice, where
a local particle distribution vector $\vec{f}$ with $q$ entries is stored on each lattice site.
Entry $f_i$ is understood to model the relative number of particles moving at a velocity $\vec{\xi}_i \in \{ -1, 0, 1 \}^d$.
We adopt writing $\bar{i}$ to refer to the population index of $f_{\bar{i}}$ moving in the direction opposite to $f_i$.
The lattice structure is specified by the common D$d$Q$q$ stencil notation.
The two- and three-dimensional first-neighborhood stencils found most prevalently in literature are
D2Q9, D3Q15, D3Q19 and D3Q27 \cite{Krueger2017}. 
Our work will also focus on these stencils.
For illustration, we employ D2Q9 throughout this paper.
D2Q9 comprises the full set of first-neighborhood velocities $\vec{\xi}_i \in \{ -1, 0, 1 \}^2$;
while their exact ordering is not relevant to our discussion, we adopt $\vec{\xi}_0 = \vec{0}$.

The lattice Boltzmann algorithm comprises the \emph{streaming} step,
where populations are advected according to their velocities;
and the \emph{collision} step, rearranging every node's populations by some collision operator $\Omega$.
Often, the algorithm also includes some type of source term $\vec{f}^F$,
which might be a momentum source modelling body forces acting on the fluid \cite{Guo2002,He1998},
or a mass source in an advection-diffusion setting \cite{Krueger2017}.
The \ac{LBM} update equations read
\begin{subequations}
    \label{eq:LbUpdateScheme}
    \begin{align}
        \vec{f}^{\ast} ( \vec{x}, t ) &= \Omega( \vec{f} (\vec{x}, t) ) + \vec{f}^F,
        \label{eq:LbCollision}
        \\
        f_{i} (\vec{x} + \vec{\xi}_i, t + \Delta t) &= f^{\ast}_{i} ( \vec{x}, t ).
        \label{eq:LbStreaming}
    \end{align}
\end{subequations}
In the remainder of this paper, we shall focus on the collision step;
the streaming step will only be mentioned briefly in \cref{section:MRTModelling}.

The original and major application of the \ac{LBM} is as 
an algorithm to simulate incompressible, Newtonian fluid flow as modelled by the \ac{NSE}.
In such simulations, the local distribution vector is found to deviate only little from a certain background distribution
\cite{Skordos1993,He1997,Geier2015,Lehmann2022}.
We denote this background distribution $\vec{f}^0$; its exact form
will be discussed in \cref{subsection:EquilibriumDistributions}.
This fact can be exploited to improve the \ac{LBM}'s numerical accuracy:
instead of storing the absolute values of the population vector,
only its deviation component $\vec{\delta f} = \vec{f} - \vec{f}^0$ can be stored per site.
As noted in \cref{section:Introduction}, we denote this as the \emph{zero-centered} storage format,
to set it apart from the classical approach, which we dub \emph{absolute} storage format.
The zero-centered storage format may be less useful in non-hydrodynamic \acp{LBM}
whenever the assumption of small deviations from the background distribution does not hold.

From the discrete distribution vector, a set of macroscopic quantities can be computed on each lattice site.
The sum over a site's populations yields the local particle density $\rho$ while their weighted mean 
becomes the macroscopic velocity $\vec{u}$, which are written as
\begin{equation}
    \label{eq:DensityAndVelocity}
    \rho := \sum_{i=0}^{q - 1} f_i, \qquad
    \vec{u} := \frac{1}{\rho} \sum_{i=0}^{q-1} f_i \vec{\xi}_i
\end{equation}
If a background distribution is given, the density can be decomposed as $\rho = \rho^0 + \delta\rho$,
with
\begin{equation}
    \label{eq:DensityAndVelocityFromDeviation}
    \rho^0 := \sum_{i=0}^{q - 1} f^0_i, \qquad
    \delta\rho := \sum_{i=0}^{q - 1} \delta f_i, \qquad
    \vec{u} := \frac{1}{\rho} \sum_{i=0}^{q-1} \delta f_i \vec{\xi}_i.
\end{equation}

\subsection{The Collision Step}
\label{subsection:UpdateEquationAlgebraic}

The site-local \ac{LBM} collision rule constitutes a relaxation process of the population vector against an equilibrium state.
In the \ac{MRT} paradigm, relaxation occurs in some collision space isomorphic to $\mathbb{R}^q$,
wherein the equilibrium state is given by a vector $\vec{q}^{\meq}$.
The population vector is transformed to and from collision space by some bijective mapping $\mathcal{T}$.
Moreover the source term may also be represented in collision space, denoted by $\vec{q}^F$ \cite{Krueger2017,Fei2017,Fei2018,Gruszczynski2020}.
This leads to the general \ac{MRT} collision equation
\begin{equation}
    \label{eq:MrtUpdateGeneral}
    \begin{split}
        \vec{q} &= \mathcal{T}(\vec{f}) \\
        \vec{f}^{\ast} &= \mathcal{T}^{-1} \left( \vec{q} + \mat{S} \left( \vec{q}^{\meq} - \vec{q} \right) + \vec{q}^F \right).
    \end{split}
\end{equation}
where $\mat{S} = \diag(\omega_0, \dots, \omega_{q-1})$ is the relaxation matrix,
specifying relaxation rates separately for each quantity $q_i$.
These quantities, and their equilibrium counterparts, usually correspond to different properties of the fluid,
and their distinct relaxation is meant to independently model different physical processes \cite{Krueger2017,Geier2015}.

To take full advantage of zero-centered storage, we aim to express \cref{eq:MrtUpdateGeneral} in deviation components only.
This is only straightforwardly possible if $\mathcal{T}$ is linear.
In this case, the decomposition of $\vec{f}$ can be reflected in the \ac{MRT} collision space,
by $\vec{q} = \vec{q}^0 + \vec{\delta q}$,
with $\vec{q}^0 = \mathcal{T} ( \vec{f}^0 )$ and $\vec{\delta q} = \mathcal{T} (\vec{\delta f})$.
Subtracting the background component, we also obtain the deviation component
$\vec{\delta q}^{\meq} = \vec{q}^{\meq} - \vec{q}^0$ of the equilibrium vector.
We thus arrive at the deviation-only update equations
\begin{equation}
    \label{eq:MrtUpdateGeneralDeviationOnly}
    \begin{split}
        \vec{\delta q} &= \mathcal{T}(\vec{\delta f}), \\
        \vec{\delta f}^{\ast} 
            &= \mathcal{T}^{-1} \left(
                    \vec{\delta q} 
                    + \mat{S} \left( \vec{\delta q}^{\meq} - \vec{\delta q} \right)
                    + \vec{q}^F
                \right).
    \end{split}
\end{equation}
By definition, a source term cannot affect the constant background component;
thus we can include $\vec{q}^F$ in the deviation-only update without alteration.

For nonlinear $\mathcal{T}$, the background-deviation-decomposition can not be transformed into collision space in this manner.
Instead, \cref{eq:MrtUpdateGeneral} is modified.
The absolute population vector is re-obtained by adding the background distribution before collision,
so that it must be subtracted again, leading to
\begin{equation}
    \label{eq:MrtUpdateAbsoluteFromZeroCentered}
    \begin{split}
        \vec{q} &= \mathcal{T}(\vec{\delta f} + \vec{f}^0), \\
        \vec{\delta f}^{\ast} 
            &= \mathcal{T}^{-1} \left(
                    \vec{q}
                    + \mat{S} \left( \vec{q}^{\meq} - \vec{q} \right)
                    + \vec{q}^F
                \right) - \vec{f}^0.
    \end{split}
\end{equation}
In the following, we will introduce the common \ac{MRT} collision spaces relevant to this work, 
and discuss the shape $\mathcal{T}$ takes therein.

\subsection{Collision Spaces}
\label{subsection:CollisionSpaces}

Assuming a D$d$Q$q$ stencil, we consider $q$-dimensional collision spaces spanned by either raw moments $m$,
central moments $\kappa$, or cumulants $C$ of the discrete population vector.
We continue to use the symbol $q$ in place of $m$, $\kappa$ and $C$ whenever we discuss general
properties of these statistical quantities.
We furthermore adopt a method of specifying statistical quantities using polynomials in the variables $x$, $y$ and $z$.
This notation is reflected exactly within the \lbmpy{} transformation system.
Given such a polynomial $p$, the respective quantity is denoted $q_p$.
For clarity, we denote \emph{monomial} quantities, defined by a monomial $x^{\alpha} y^{\beta} z^{\gamma}$, as $q_{\abc}$.
A collision space based on a specific type of quantity is now defined through its \emph{basis},
where a basis is a sequence $\left(p_i\right)_{i = 0, \dots, q - 1}$ of linearly independent polynomials.

\subsubsection{Raw Moments}

Originally, the \ac{MRT} methods based on raw moments were developed to overcome
deficiencies
of the discrete \ac{BGK} operator \cite{Krueger2017,dHumieres2002MRT}.
Among their advantages are the possibility to separate shear from bulk viscosity,
and the ability to tune higher-order relaxation rates in order to improve
stability and accuracy \cite{Krueger2017}.

The raw moment-generating function $M$ of a distribution $f$ is defined using the multidimensional
Laplace transform $\mathcal{L}$ as
\begin{equation}
    \label{eq:MomentGeneratingFunction}
    M(\vec{\Xi}) := \mathcal{L} \{ f \} (- \vec{\Xi}).
\end{equation}
Since $\mathcal{L}$ is an integral operator, $f$ must be integrable.
Here the discrete distribution vector is made integrable by means of the 
Dirac delta distribution as $f(\vec{\xi}) := \sum_i f_i \delta(\vec{\xi} - \vec{\xi}_i)$.
The monomial raw moments $m_{\abc}$ of $f$ are now defined as the mixed derivatives of $M$,
evaluated at zero
\begin{equation}
    \label{eq:RawMomentsFromGenFunc}
    m_{\abc} :=
    \left. \partial_1^{\alpha} \partial_2^{\beta} \partial_3^{\gamma} 
                    M( \vec{\Xi} )
                \right\vert_{\vec{\Xi} = \vec{0}} .
\end{equation}
In the discrete case, this evaluates to a simple summation over the entries of the population vector.
Polynomial raw moments $m_p$ are obtained as linear combinations of their monomial components.
This linear combination can be combined with the sum obtained from the generating function.
Together, monomial and polynomial quantities may be computed as
\begin{equation}
    \label{eq:DiscreteRawMomentsDef}
    m_{\abc} = \sum_{i} f_i \xi_{i,x}^{\alpha} \xi_{i,y}^{\beta} \xi_{i,z}^{\gamma}, \quad
    m_{p} = \sum_{i} f_i p( \vec{\xi}_i ).
\end{equation}

For a given basis $\left( p_i \right)_i$, the transformation to discrete raw moment space can now be represented
as an invertible matrix $\mat{M}$.
Thus, $\vec{m}_p = \mathcal{T} (\vec{f}) = \mat{M} \vec{f}$ is a linear mapping,
allowing us to reflect the background-deviation-decomposition in raw moment space.

\subsubsection{Central Moments}

\acp{LBM} based on raw moments violate Galilean invariance. 
To correct for this, \ac{MRT} methods based on relaxing central moments of the discrete distribution function
have been developed \cite{Geier2006,Geier2006Diss,Premnath2011CM3D,Coreixas2019}.
In contrast to raw moments, central moments are taken with respect to the co-moving frame
of reference, given by the macroscopic velocity $\vec{u}$.

The velocity shift is reflected in Laplace frequency space by multiplication
with $\exp \left( - \vec{\Xi} \cdot \vec{u} \right)$,
which we use to define the central moment-generating function
\begin{equation}
    \label{eq:CentralMomentGeneratingFunction}
    K(\vec{\Xi}) := \exp \left( - \vec{\Xi} \cdot \vec{u} \right) M(\vec{\Xi}).
\end{equation}
Its derivatives, in turn, give rise to the monomial central moments $\kappa_{\abc}$,
from which polynomial central moments $\kappa_p$ are again obtained as linear combinations.
The shifted frame of reference appears again in the explicit equations via subtraction
\begin{equation}
    \label{eq:DiscreteRawMomentsDef}
    \kappa_{\abc} = \sum_{i} f_i 
        (\xi_{i,x} - u_x )^{\alpha}
        (\xi_{i,y} - u_y )^{\beta}
        (\xi_{i,z} - u_z )^{\gamma}, \quad
    \kappa_{p} = \sum_{i} f_i p( \vec{\xi}_i - \vec{u} ).
\end{equation}
These transform equations are again linear, giving rise to the central moment matrix $\mat{K}$
which defines $\mathcal{T}$ for the basis $\left( p_i \right)_i$.

\subsubsection{Cumulants}
\label{subsubsection:Cumulants}

Cumulants, other than any kind of moment, characterize a distribution's shape independent of any frame of reference \cite{Geier2015}.
They thus share the velocity-independence of central moments, but additionally introduce
mutual statistical independence.
Similar to moments, monomial cumulants $c_{\abc}$ are the mixed derivatives of the cumulant-generating function at the origin.
The cumulant-generating function is defined as the natural logarithm of the moment-generating function
\begin{equation}
    \label{eq:CumulantGeneratingFunction}
    C(\vec{\Xi}) := \log M(\vec{\Xi}),
    \quad
    c_{\abc} := \left. \partial_1^{\alpha} \partial_2^{\beta} \partial_3^{\gamma} 
                    C( \vec{\Xi} )
                \right\vert_{\vec{\Xi} = \vec{0}},
    \quad
    C_{\abc} := \rho c_{\abc}.
\end{equation}
We follow Geier et al.\ \cite{Geier2015} in defining the rescaled cumulants $C_{\abc}$.
We give no explicit equations for monomial or polynomial cumulants in terms of populations;
instead, we discuss a more practical approach to their computation in \cref{section:DerivationAndCodegen}.
Due to the cumulant transform's nonlinearity, no reflection of the population's decomposition
in cumulant space is possible.

\subsection{Equilibrium and Background Distributions}
\label{subsection:EquilibriumDistributions}

\begin{figure}
    \begin{center}
        \includegraphics{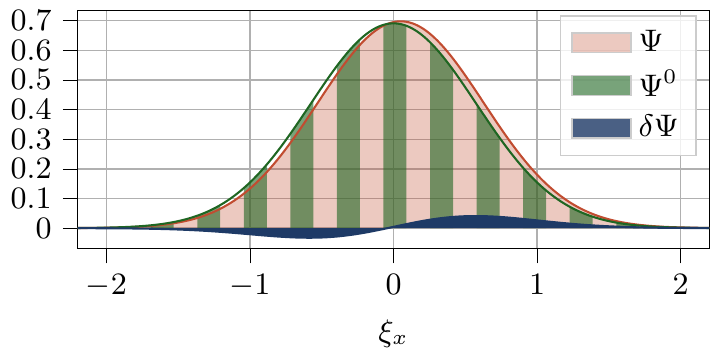}
    \end{center}
    \caption{Plot of $\Psi$, $\Psi^0$ and $\delta\Psi$ in one dimension at $\rho = 1.01$, $u_x = 0.05$.}
    \label{fig:1DMaxwellianGraph}
\end{figure}

The basis of most \acp{LBM} is a variant of the Maxwell-Boltzmann distribution \cite{Krueger2017}
that defines the equilibrium.
Depending on density $\rho$, velocity $\vec{u}$, and the speed of sound parameter $c_s$,
its continuous form reads
\begin{equation}
    \label{eq:ContMaxwellian}
    \Psi \left( \rho, \vec{u}, \vec{\xi} \right)
    = \rho  \left(
                \frac{1}{2 \pi c_s^2}
            \right)^{3/2}
            \exp \left(
                - \frac{
                    \left\Vert
                        \vec{\xi} - \vec{u}
                    \right\Vert^2
                }{2 c_s^2}
            \right).
\end{equation}
Since, in hydrodynamic simulations, the fluid density deviates only little from a background density $\rho^0$,
we take as a background distribution the fluid rest state at $\rho = \rho^0, \vec{u} = \vec{0}$.
We hence denote $\Psi^0 = \Psi(\rho^0, \vec{0})$
and define the deviation component of the equilibrium distribution as $\delta\Psi = \Psi - \Psi^0$.
The background density is typically set to unity.
\Cref{fig:1DMaxwellianGraph} illustrates the relation between $\Psi$, its background and deviation component,
with density and velocity chosen from the typical simulation range.
We observe a large difference in magnitude between background and deviation component;
overall, the background component dominates the distribution.

Representations of $\Psi$ and $\Psi^0$ in raw moment, central moment, and cumulant space
can be obtained by algebraically computing and differentiating the respective generating functions.
In case of raw and central moments, this reduces to the well-known moment integrals.
The deviation component $\delta \Psi$ has representations in both moment spaces, but not in cumulant space,
since the cumulants of $\delta\Psi$ are undefined at the singularity $\rho = \rho^0$,
where its area under the curve is zero.
This can be verified by expanding cumulants in terms of raw moments:
Applying the chain rule to evaluate \cref{eq:CumulantGeneratingFunction}
produces equations containing divisions by $\delta\rho$.

Given e.g. a vector $m^{0}$ of $q$ independent raw moments of $\Psi^0$,
the background distribution in population space may be obtained as $\vec{f}^0 = \mat{M}^{-1} \vec{m}^0$.
In the case of the standard hydrodynamic equilibrium, this vector reduces to the lattice weights $\vec{w}$.

Alternatively the equilibrium state may also be specified as a discrete distribution $\vec{f}^{\meq}$.
In this case, its representation in collision space is simply obtained as
$\vec{q}^{\meq} = \mathcal{T} \left( \vec{f}^{\meq} \right)$.

    \section{MRT Method Modelling Frontend}
\label{section:MRTModelling}

The main component our code generation system is an automatic procedure that takes an abstract
\ac{LBM} specification and derives from it a sequence of symbolic equations
implementing one of the collision rules \cref{eq:MrtUpdateGeneral,eq:MrtUpdateGeneralDeviationOnly,eq:MrtUpdateAbsoluteFromZeroCentered}.
This specification is formulated using a flexible Python \ac{API}.
Both the modelling \ac{API} and the derivation system itself rely on the computer algebra package \SymPy{} \cite{sympy}
to represent and manipulate all components of an \ac{LBM} in symbolic, mathematical form.
The automatic derivation and thus the software performing these transformations follows closely the theory of \cref{section:TheoreticalBackground}.
This section introduces the user front-end for modeling,
while the next section will focus on the derivation and code generation procedures.

\begin{figure}
    \begin{center}
        \includegraphics{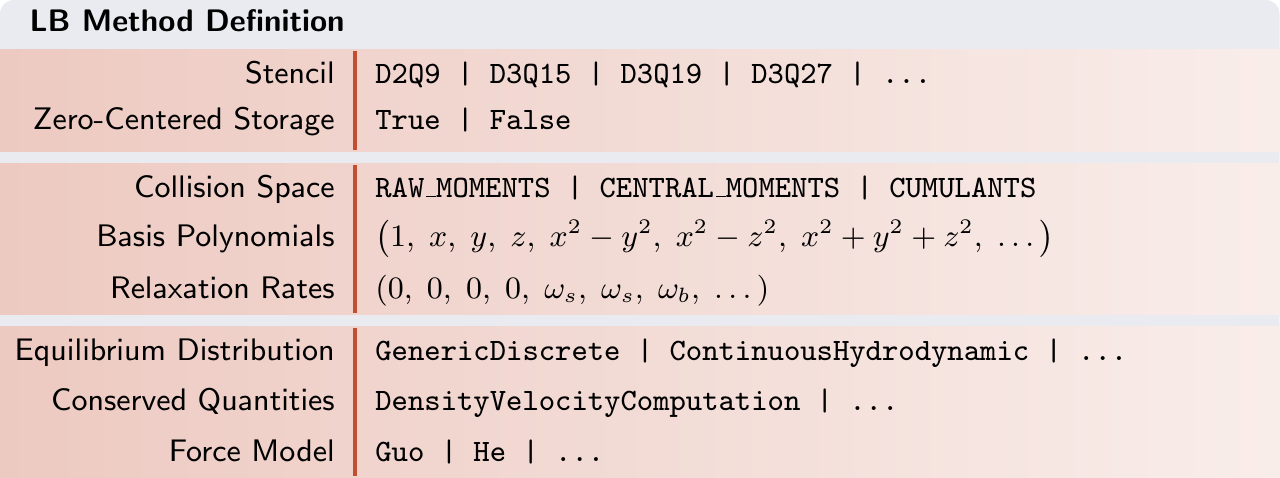}
    \end{center}
    \caption{
        Configuration space for modelling \ac{MRT}-type methods in \lbmpy{}.
        Method parameters are split into three groups.
        The first group governs structure and storage format of the discrete lattice,
        the second group describes the method's collision space,
        and the third group comprises those components governing the actual collision process.}
    \label{fig:LbMethodModelling}
\end{figure}

The parameter space of \lbmpy{}'s abstract method specification is shown in \cref{fig:LbMethodModelling}.
The lattice structure is defined by selecting a stencil and a storage format (absolute or zero-centered).
Next, the collision space is specified by fixing one type of statistical quantity
and stating its basis as a sequence of polynomials.
Additionally, corresponding relaxation rates must be specified.
Each relaxation rate can be given either as a fixed numerical value
or in symbolic form, allowing its value to remain undetermined until runtime of the generated code.

The definition's final three components require a significantly more elaborate structure.
To model equilibrium distributions, the computation of macroscopic quantities, and force models,
\lbmpy{} provides specific hierarchies of Python classes.
Abstract base classes define their respective interfaces.
These components do not only encapsulate information about the method,
but also take an active role during the code generation process.

For instance, the equilibrium object must not only produce an algebraic form (discrete or continuous) of its distribution;
it must also distinguish between absolute form and delta-equilibrium and provide the background distribution in the latter case.
Furthermore, it must provide methods to compute its raw moments, central moments, and cumulants.
These methods are invoked during the symbolic derivation of the collision rule.
The same holds for force models and conserved quantity computation; both components will provide parts of the equations
making up the collision rule later on.

Apart from functional requirements, there is another significant advantage to using classes.
While certain components are already implemented in the current version of \lbmpy{},
such as the Maxwellian equilibrium, the Guo \cite{Guo2002} and He \cite{He1998} force models, etc.,
this structure makes \lbmpy{} flexible and extensible.
In particular, a developer using \lbmpy{} may use the interfaces of the respective base classes in order to implement
arbitrary custom equilibrium distributions, force models, or computation procedures for the macroscopic quantities.
An example of this will be shown in \cref{subsection:ShallowWaterApplication}.

The user may freely combine various options for different method parameters, with some restrictions.
For instance, deviation-only equilibrium instances may only be used in junction with zero-centered storage.
Furthermore, as already discussed in previous sections, cumulant space is incompatible with deviation-only equilibria.
As we will demonstrate in \cref{subsection:ShallowWaterApplication},
this interface is versatile and flexible both for implementing common hydrodynamic \acp{LBM},
and also for rapidly constructing more specialized methods that are not natively
realized within \lbmpy{}.

    \section{Derivation Procedure and Code Generation}
\label{section:DerivationAndCodegen}

The front-end \ac{API} described in the previous
section can be used to specifiy a concrete instance of an \acs{LB} method.
This specification serves as input 
to our automatic derivation and code generation pipeline.
This pipeline comprises several steps, ultimately emitting an optimized numerical kernel
implementing the collision rule for either CPU or \ac{GPU} architectures.
It relies on the domain-specific language of our code generation framework \pystencils{} \cite{Bauer2019pystencils},
which in turn is based on \SymPy{}.
In the following, we shall describe the stages of this pipeline as implemented in
version {1.1} of \lbmpy{}\footnote{\href{https://i10git.cs.fau.de/pycodegen/lbmpy/-/tree/release/1.1}{i10git.cs.fau.de/pycodegen/lbmpy/-/tree/release/1.1}}.

\subsection{The Collision Rule}

The first stage of the pipeline takes the abstract definition to derive the equations constituting
the collision rule.
Depending on the combination of storage and equilibrium format, an implementation of either
\cref{eq:MrtUpdateGeneral,eq:MrtUpdateGeneralDeviationOnly,eq:MrtUpdateAbsoluteFromZeroCentered}
is derived in symbolic form.
Manipulating the collision equations on the mathematical level, we can leverage all information available
about the method to simplify and optimize the equations.
The derivation system is modular, combining equations generated by several components.
Among these are the equilibrium and force model instances,
providing algebraic expressions of their respective representations in the given collision space;
as well as the conserved quantity computation, which produces equations for density (or its analogues)
and velocity (cf.\ \cref{eq:DensityAndVelocity,eq:DensityAndVelocityFromDeviation}).

The equations forming the collision space transformation $\mathcal{T}$
are provided by a set of dedicated classes within \lbmpy{}.
Apart from the simple symbolic matrix-vector-multiplication,
\lbmpy{} provides two types of highly efficient Chimera-based transforms for raw and central moment space,
which we will discuss in the following subsections.

The equations produced by these various components are combined to produce the collision rule
as a sequence of assignments of mathematical expressions to symbolic variables.
This assignment collection is passed on to \lbmpy{}'s sophisticated simplification procedure,
described in \cref{subsection:Simplification},
and will ultimately be transformed into a numerical kernel (cf.\ \cref{subsection:StreamingAndUpdate,subsection:CodeGeneration}).

\subsubsection{Raw Moment Transform}

To compute raw moments from pre-collision populations, we first implement a Chimera transform to obtain
monomial raw moments, which are afterward recombined to polynomials.
Chimera transforms were first introduced in \cite{Geier2015} for transforming to central moment space.
Using intermediate \enquote*{chimera} quantities $m_{x|\beta \gamma}$ and $m_{xy | \gamma}$,
our raw-moment Chimera transform reads
\begin{equation}
    \label{eq:RawMomentChimeraTransform}
    \begin{split}
        f_{xyz} &:= \begin{cases}
            f_i & \text{if } \vec{\xi}_i = (x,y,z)^T \text{ is contained in stencil} \\
            0 & \text{otherwise}
        \end{cases} \\
        m_{xy|\gamma} &:= \sum_{z \in \{-1, 0, 1\} } f_{xyz} \cdot z^{\gamma} \\
        m_{x|\beta \gamma} &:= \sum_{y \in \{-1, 0, 1\}} m_{xy|\gamma} \cdot y^{\beta} \\
        m_{\alpha \beta \gamma} &:= \sum_{x \in \{-1, 0, 1\}} m_{x|\beta \gamma} \cdot x^{\alpha}.
    \end{split}
\end{equation}
Due to its recursive nature, the Chimera transform minimizes the number of arithmetic operations,
since each possible combination of populations and velocities is evaluated exactly once. 

To transform from raw moments back to populations, we first decompose the polynomial moments into their monomial components.
We then derive $\vec{f}^{\ast} = \mat{M}^{-1} \vec{m}^{\ast}$ by symbolic matrix-vector-multiplication.
Those equations we simplify by splitting the expressions for populations $f_i$ and $f_{\bar{i}}$
moving in opposite directions into their symmetric and antisymmetric parts, such that
\begin{equation}
    \label{eq:RawMomentChimeraBackwardSymmetric}
    f_i^{\ast} = f_i^{\mathrm{+}} + f_i^{\mathrm{-}}, \qquad
    f_{\bar{i}}^{\ast} = f_i^{\mathrm{+}} - f_i^{\mathrm{-}}.
\end{equation}
This split roughly cuts the number of arithmetic operations in half.

\subsection{Central Moment Transform}

In \lbmpy{}, central moments are not transformed from and to populations directly,
but using monomial raw moments as intermediates.
To find an efficient transform between monomial raw and central moments, we observe that they
are bidirectionally related through binomial expansions
\begin{subequations}
    \begin{align}
        \label{eq:RawToCentralMomentsBinomial}
        \kappa_{\abc} &= 
            \sum_{a, b, c = 0}^{\alpha, \beta, \gamma}
            \binom{\alpha}{a} \binom{\beta}{b} \binom{\gamma}{c}
                (- u_x)^{\alpha - a} (- u_y)^{\beta - b} (- u_z)^{\gamma - c}
                m_{abc}
        , \\
        \label{eq:CentralToRawMomentsBinomial}
        m_{\abc} &= 
            \sum_{a, b, c = 0}^{\alpha, \beta, \gamma}
            \binom{\alpha}{a} \binom{\beta}{b} \binom{\gamma}{c}
                u_x^{\alpha - a} u_y^{\beta - b} u_z^{\gamma - c}
                \kappa_{abc}.
    \end{align}
\end{subequations}
In both directions we may separate the nested sums, introducing them as Chimera quantities.
This leads us to the binomial Chimera transforms between raw and central moments,
which we state
\begin{align}
    \kappa_{ab|\gamma}  &:=
        \sum_{c = 0}^{\gamma} \binom{\gamma}{c} (- u_z)^{\gamma - c} m_{abc} &
    m^{\ast}_{ab|\gamma}  &:=
        \sum_{c = 0}^{\gamma} \binom{\gamma}{c} u_z^{\gamma - c} \kappa^{\ast}_{abc} \notag \\
    \kappa_{a|\beta\gamma} &:=
        \sum_{b = 0}^{\beta} \binom{\beta}{b} (- u_y)^{\beta - b} \kappa_{ab|\gamma} &
    m^{\ast}_{a|\beta\gamma} &:=
        \sum_{b = 0}^{\beta} \binom{\beta}{b} u_y^{\beta - b} m^{\ast}_{ab|\gamma} \\
    \kappa_{\alpha\beta\gamma} &:=
        \sum_{a = 0}^{\alpha} \binom{\alpha}{a} (- u_x)^{\alpha - a} \kappa_{a|\beta\gamma} &
    m^{\ast}_{\alpha\beta\gamma} &:=
        \sum_{a = 0}^{\alpha} \binom{\alpha}{a} u_x^{\alpha - a} m^{\ast}_{a|\beta\gamma}  \notag
\end{align}
Again, as every combination of moments and velocities is evaluated exactly once,
this results in expressions that require only a minimal number of arithmetic operations.
Polynomial central moments are combined from monomial quantities after the forward transform
and decomposed before the backward transform.

\subsubsection{Cumulant Transform}

Similar to their original formulation in \cite{Geier2015}, cumulants in \lbmpy{} are obtained
from central moments. 
To derive their nonlinear relation, we re-express the cumulant-generating function \cref{eq:CumulantGeneratingFunction}
in terms of the central moment-generating function, to obtain the bidirectional relations
\begin{equation}
    \label{eq:CumulantAndCentralMomentGenFuncs}
    C \left( \vec{\Xi} \right)
        = \left( \vec{\Xi} \cdot \vec{u} \right) + \log K \left( \vec{\Xi} \right)
    \quad \Leftrightarrow \quad
    K( \vec{\Xi} ) = \exp \left( C(\vec{\Xi}) - \vec{\Xi} \cdot \vec{u} \right).
\end{equation}
We then employ \SymPy{}'s symbolic differentiation capabilities to obtain expressions for the derivatives of $C$
in terms of derivatives of $K$, and vice versa.
Substituting monomial central moments and cumulants for these derivatives,
we arrive at the equations for both transform directions.
It must be noted that the equations obtained this way contain logarithms and exponential functions,
whose presence is undesirable in a numerical kernel.
However, they are only associated with the zeroth-order cumulant $C_{000}$,
allowing us to eliminate them in a global simplification step.

\subsection{A Posteriori Simplifications}
\label{subsection:Simplification}

Once the derivation of the collision kernel is complete,
there will still be several possibilities to simplify and optimize the resulting kernel by symbolic manipulations.
Our simplifiation procedure in \lbmpy{} combines generic algebraic simplifiations with optimizations
designed specifically for \acp{LBM}, utilizing information about the method available
during code generation.
In particular, \lbmpy{} applies the following simplification steps to the generated sequence of
assignments.
For advanced users of \lbmpy{} it is additionally possible to disable these transformations or to customize them. 
\begin{description}
    \item[Conserved Quantity Rewriting] Equations computing $\rho$, $\vec{u}$, etc.\ from populations directly are
            replaced by equations involving zeroth- and first-order raw moments.
    \item[Collapsing Conserved Central Moments] We collapse the equations for zeroth- and first-order central moments,
            e.g.\ obtaining, $\kappa_{000} = \rho$ and $\kappa_{100} = - F_x / 2$.
    \item[Propagation of Logarithms] To simplify logarithmic and exponential expressions in cumulant-based collision rules,
            we propagate certain assignments containing logarithms to their usages;
            thus canceling them with the corresponding exponential functions.
    \item[Common Subexpression Elimination] We use \SymPy{}'s \ac{CSE} feature to extract common
            terms into separate assignments, as previously described in \cite{Bauer2021lbmpy}.
    \item[Expression Propagation] Assignments whose right-hand sides are constant, single symbols,
            products of macroscopic quantities, or multiples of body force components, are propagated to their usages.
    \item[Unused Subexpression Elimination] A simple dependency analysis based on the \ac{SSA} form allows us to
            eliminate any assignments whose left-hand side symbols are never used.
\end{description}

While all of these simplifications can help in reducing kernel complexity,
we found the combination of the final two to have the largest impact.
Their strength hinges on the fact that \SymPy{}, as a computer algebra system,
automatically evaluates any constant terms, and attempts
to cancel equal but opposite terms in any constructed expression. 
Our propagation steps trigger that behaviour automatically.
Apart from minor eliminations occuring throughout the equations, this has two major effects.

First, quantities that are invariant under relaxation are propagated through the relaxation step, and inserted directly into the
backward transform equations. This, in turn, leads to an automatic simplification of these equations.
Apart from reducing overall operation count, this step also serves to eliminate algebraically idempotent operations.
To given an example: Without this propagation, relaxation and backward binomial Chimera transform would include assignments similar to
\begin{equation}
    \begin{split}
        \kappa_{000} &= \rho \\
        \kappa_{100} &= - F_x / 2 \\
        \kappa^{\ast}_{000} &= \kappa_{000} \\
        \kappa^{\ast}_{100} &= \kappa_{100} + F_x \\
        m^{\ast}_{10|0} &= \kappa^{\ast}_{000} u_x + \kappa^{\ast}_{100}.
    \end{split}
\end{equation}
The propagation steps described above will cause most of these assignments to be dropped, leaving us with just
\begin{equation}
    m^{\ast}_{10|0} = \rho u_x + F_x / 2.
\end{equation}

The second and more significant effect of alias and constant propagation occurs whenever 
some relaxation rates are set to unity.
In this case, propagation enables the elimination of large portions of the forward collision space transform,
as well as significant algebraic simplification of the backward transform.
Substituting $\omega = 1$, a relaxation equation
$q^{\ast}_p = q_p + \omega \left( q^{\meq}_p - q_p \right)$
is immediately simplified to $q^{\ast}_p = q^{\meq}_p$.
The forward transform equations for $q_p$ are then no longer required, allowing us to drop them.
This has the potential to massively decrease the overall operation count, since the transform equations especially
for higher-order quantities constitute a significant portion of the collision kernel.
This will be discussed in further detail below.

If $q^{\meq}_p$ is a simple expression, it can be propagated into the backward transform equations.
The effectiveness of this process is improved by the fact that many statistical quantities
of common equilibrium distributions are zero.
Propagating a zero into the backward transform equations will typically eliminate only a small number of summands; in fact,
for the backward Chimera transform, only a single term is removed from one innermost Chimera.
Still, the more quantities are relaxed with $\omega_i = 1$, the more significant this effect will grow.
Propagation of zeroes is most effective with the cumulant backward transform, due to its nonlinearity.
The equations computing higher-order post-collision central moments from post-collision cumulants are by far the most complex
derived anywhere within our framework. 
They do, however, contain a significant number of products of post-collision cumulants.
Substituting those cumulants' equilibrium values, which are mostly equal to zero, several such multiplications may be eliminated.
\begin{table}
    \caption{%
        Arithmetic operation counts in the symbolic representation of compute kernels
        generated by \lbmpy{} for several method definitions on the D2Q9, D3Q19 and D3Q27 stencils.
        Numbers for kernels derived without simplification (N), standard simplification (S) and
        simplification plus \ac{CSE} (S+CSE) are listed.
    }
    \begin{center}
        \includegraphics{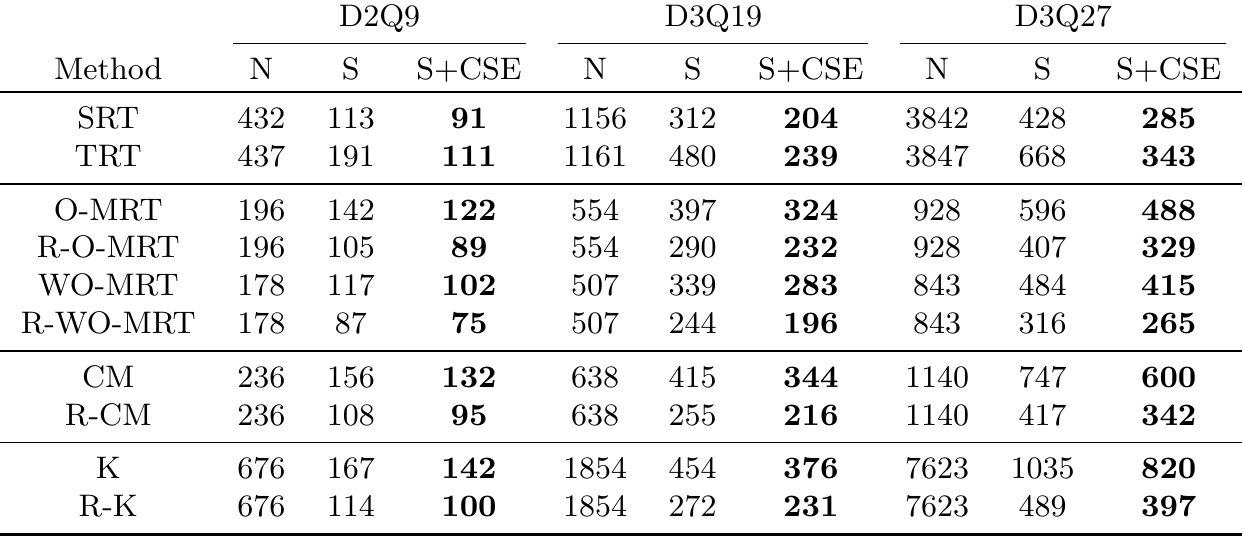}
    \end{center}
    \label{table:MethodOperationCounts}
\end{table}

To illustrate the efficacy of our simplification strategy, we count the total number of operations in the collision kernels
generated by \lbmpy{} from a number of different method definitions.
All methods employ zero-centered storage and the default continuous Maxwellian equilibrium, whose moments are truncated at
the second order in velocity.
No force model is employed.
Kernels are derived for non-weighted orthogonal \ac{MRT} (denoted O-MRT)
and weighted orthogonal \ac{MRT} (WO-MRT) in raw moment space \cite{Krueger2017};
as well as polynomial central moment (CM) \cite{Geier2006,Rosis2016,Rosis2017}
and polynomial cumulant (K) \cite{Geier2015} methods.
We compare these with implementations of the \ac{SRT} and \ac{TRT} collision operators,
also derived by \lbmpy{} using the original procedure published in \cite{Bauer2021lbmpy}.
The prefix \enquote*{R-} denotes a regularized method; all but the relaxation rates governing shear viscosity are therein set to unity.
Such regularization is permissible since higher-order \enquote*{ghost} modes do not affect simulated physics \cite{Geier2015,Geier2017};
further, it is a common measure to improve a method's stability \cite{Latt2005,Geier2020,Coreixas2019}.
Otherwise, every relaxation rate is represented by a dedicated symbol $\omega_i\, (i = 0, \dots, q-1)$.
The population and raw moment space methods use the delta-equilibrium for relaxation, while the absolute equilibrium is used in central
moment and cumulant space.

\Cref{table:MethodOperationCounts} lists the arithmetic operation counts of kernels generated for those methods, with three different
levels of simplification intensity: None at all; full simplification without \ac{CSE}; and full simplification including \ac{CSE}.
The table shows that \lbmpy{} is capable of producing kernel implementations for complex and advanced
lattice Boltzmann methods with only little overhead when compared to simple \ac{SRT} or \ac{TRT} kernels.
In fact, the regularized WO-MRT kernels are even visibly smaller in size than the basic \ac{SRT} kernels.
This observation holds also in comparison with carefully hand-crafted implementations,
as in \cite{Wellein2006}, where Wellein et al.\ report about 200 operations for the D3Q19 \ac{SRT} kernel.
Note that, for non-regularized raw moment \ac{MRT}, we observe vast improvements (almost 50 \% fewer operations)
compared to the results originally published for \lbmpy{} in \cite{Bauer2021lbmpy}.
Furthermore, considering the cumulant-based \ac{LBM}, we surpass the carefully optimized implementation of Geier et al.,
who in \cite{Geier2020} report 432 arithmetic operations just for the forward transform.
Note, however, that these numbers describe implementations in a high-level programming language or, in our case, in symbolic form.
The actual number of floating-point operations in the compiled code may differ due to transformations applied by modern
optimizing compilers.

\begin{table}
    \caption{Savings in arithmetic operations due to regularization for several methods on the D3Q27 stencil.
        We display the savings relative to the operation count of the kernel variant with purely symbolic relaxation rates.
        \enquote*{Higher Order Regularized} implies that only the relaxation rates for statistical quantites of order 5 and 6
        are set to unity. \enquote*{Fully Regularized} methods have all but their shear relaxation rates set to one.
        All kernels were simplified as far as possible, without applying \ac{CSE}.}
    \begin{center}
        \includegraphics{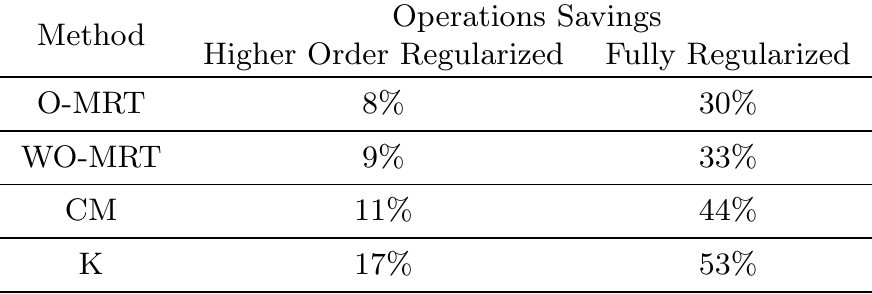}
    \end{center}
    \label{table:D3Q27RegularizationSavings}
\end{table}

Comparing the kernel sizes of methods using purely symbolic relaxation rates with 
their regularized variants, the effectiveness of our expression propagation steps becomes apparent.
\Cref{table:D3Q27RegularizationSavings} lists the savings in operation counts with respect to the non-regularized version for
orthogonal raw-moment, central moment and cumulant methods on the D3Q27 stencil.
We compare both the fully regularized methods, whose operation counts are shown in \cref{table:MethodOperationCounts},
and only partially (higher-order) regularized methods, with only the relaxation rates for quantities of order $\ge 5$ set to unity.
The reduction in operation count achieved through regularization grows with the complexity of the collision equations.
The largest part of these savings comes from the elimination of transform equations.
In the central moment method's kernel, the forward transform equations for the fifth and sixth order moments alone make for
roughly eight percent of the total operation count, which can simply be omitted with regularization.
The largest potential for simplifiation exists for cumulant-based methods:
on D3Q27, ten percent of arithmetic operations can be attributed to just the four equations
computing the monomial pre-collision cumulants $C_{122}$, $C_{212}$, $C_{221}$ and $C_{222}$ from central moments.

\subsection{Streaming Patterns and Update Rule}
\label{subsection:StreamingAndUpdate}

The collision rule generated by stage one constitutes the relaxation process for a single cell,
whose populations are represented purely symbolically.
In the next stage, we employ the field abstraction of \pystencils{} to map the rule over the cells of a $d$-dimensional
field data structure, holding $q$ values per cell.
Therein, the iteration over the field is abstracted through \emph{field accesses}; special kinds of symbols that model
accesses relative to the current cell.
A field access reading \texttt{f[x, y, z](i)} corresponds to the $i$-th entry of the local cell's neighbor
with integer offset $(x, y, z)$.
The correspondence between pre- and post-collision populations, and relative field accesses, is governed by the
\emph{streaming pattern}.
At the time of writing, \lbmpy{} supports the classical pull- and push-patterns, which require separate arrays
for reading and writing to avoid data conflicts \cite{Wittmann2013};
as well as four different in-place streaming patterns. 
With these techniques the double data structures can be avoided.
They are the AA-pattern \cite{Bailey2009}, Esoteric Twist \cite{Geier2017EsoTwist},
Esoteric Pull, and Esoteric Push \cite{Lehmann2022Streaming}.
The field update rule is constructed from the collision rule by substituting field accesses for symbolic populations,
according to the specified streaming pattern.

\subsection{Code Generation}
\label{subsection:CodeGeneration}

As a last step, the update rule is passed on to the \pystencils{} code generator \cite{Bauer2019pystencils}.
It transforms its symbolic equations into compilable C or CUDA code; wrapping it either with nested loops
over the field, or prepending GPU indexing code.
Furthermore, the code generation system may optionally apply acceleration techniques such as
OpenMP parallelization \cite{openmp2021},
loop splitting, or loop blocking \cite{Bauer2019pystencils}.
Finally, the generated kernel can be compiled and loaded directly within the Python environment,
or it can be written to a file to be integrated into separate code bases.
The former option allows for rapid development and testing of \ac{LB} methods,
while the latter enables integration with large-scale simulation frameworks,
like \waLBerla{} \cite{Bauer2021lbmpy,Bauer2021waLBerla,Holzer2021}.

    \section{Applications}
\label{section:Applications}

In this section, we present a numerical benchmark involving Taylor-Green vortex decay to assess
the impact of zero-centered storage. 
As a second example we employ \lbmpy{} to generate an \ac{LBM} for the \acl{SWE}
to illustrate its extensibility to novel application fields.

\subsection{Taylor Green Vortex}

In order to assess the effectivenes of our generalization of zero-centered storage to moment spaces
in improving numerical accuracy and reducing round-off error,
we replicate a test case recently published in \cite{Lehmann2022} involving the Taylor-Green vortex flow.
Therein, a periodic box of vortices with velocity magnitude $u_0$ is initialized,
their transient decay simulated,
and compared to the known analytic solution.
The analytic solution, which also specifies the initial flow field, reads
\begin{equation}
	\label{eq:TGA_init}
	\begin{split}
	u_x(t) &= u_0 \, \cos\left(\kappa \, x\right) \, \sin\left(\kappa \, y\right) \exp\left(-2 \nu \kappa^2 t\right) \\
	u_y(t) &= - u_0 \, \sin\left(\kappa \, x\right) \, \cos\left(\kappa \, y\right) \exp\left(-2 \nu \kappa^2 t\right) \\
	u_z(t) &= 0 \\
	\rho \left(t\right) &= 1 - \frac{3 \, u_0^2}{4} \left(\cos\left(2 \,\kappa \, x\right) + \cos\left(2 \, \kappa \, y\right)\right) \exp\left(-4 \nu \kappa^2 t\right)
	\end{split}
\end{equation}

\begin{figure}[t]
	\begin{center}
		\includegraphics[trim={0 0 0 100}, clip, width=0.8\linewidth]{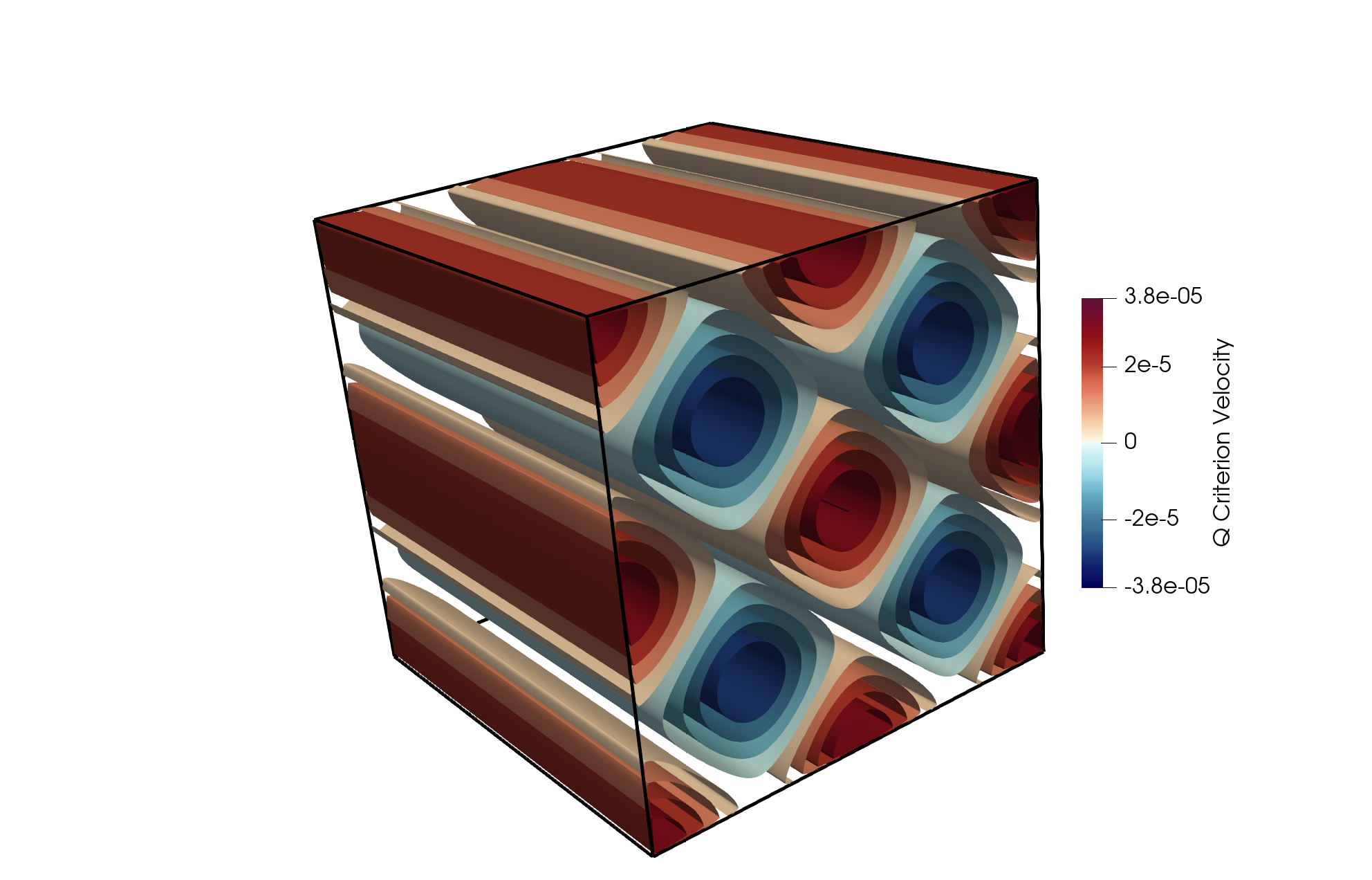}
	\end{center}
	\caption{Initialization of the Taylor-Green vortex test case with $u_0 = 0.25$, $\nu = \frac{1}{6}$, $\omega = 1$, $\kappa = \frac{2 \pi}{L}$ and $L = 256$.}
	\label{fig:InitTGA}
\end{figure}

The system is initialized at $t = 0$ with $u_0 = 0.25$.
We set the kinematic shear viscosity to $\nu = \frac{1}{6}$, which results in the viscosity-governing relaxation rate $\omega = 1$.
Furthermore, $\kappa = \frac{2 \pi}{L}$ and $L = 256$ is the side length of the squared domain.
The initial state of the flow field is displayed in \cref{fig:InitTGA}.

The kinetic energy is calculated as
\begin{equation}
	\label{eq:TGA_kin_energy}
	E \left(t\right) 
		= \int_{0}^{L} \int_{0}^{L}  \int_{0}^{L}  \frac{\rho}{2} \left(u_x^2 + u_y^2 + u_z^2\right) \dd x \, \dd y \, \dd z
		= u_0^2 \pi^2 \exp\left(-4 \nu \kappa^2 t\right),
\end{equation}
and we denote the initial kinetic energy as $E_0 = E\left( 0 \right)$.

\begin{figure}[t]
	\begin{center}
		\includegraphics{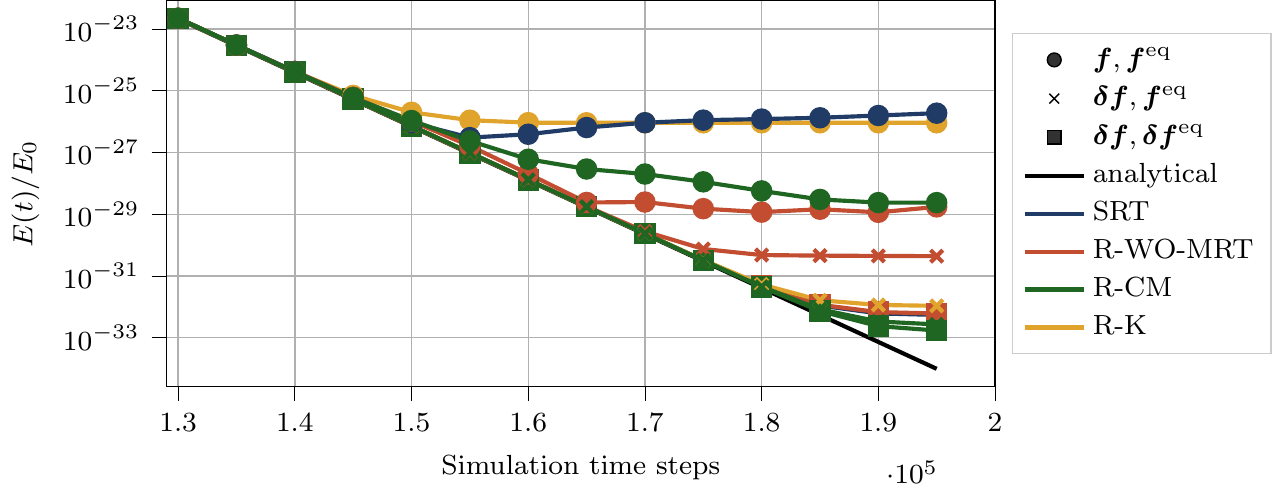}
	\end{center}
	\caption{%
		Relativ energy $E\left(t\right) / E_0$ for various D3Q27 LB methods.
		The PDFs are either stored in the absolute (circular marker) or zero-centered format,
		the latter relaxed against the absolute (x marker) or delta-equilibrium (square marker).
		We show only a range of timesteps where deviation from the analytical solution occurs.
	}
	\label{fig:TGAPlot}
\end{figure}
\begin{table}[t]
	\label{tab:numericTGA}
	\caption{%
		Relativ energy $E\left(t\right) / E\left(E_0\right)$  for various D3Q27 LB methods after \num{200000} timesteps
		for all admissible combinations of storage and equilibrium format.
	}
	\begin{center}
		\begin{tabular}{cccc}
			\multirow{2}{*}{Collision space} & \multirow{2}{*}{absolute storage} & \multicolumn{2}{c}{zero-centered} \\
			&           & $\vec{f}^{\meq}$ & $\vec{\delta f}^{\meq}$ \\
			\toprule
			
			SRT & $1.9 \cdot 10^{-26}$ & - & $5.4 \cdot 10^{-33}$ \\
			R-WO-MRT & $1.7 \cdot 10^{-29}$ & $4.4 \cdot 10^{-31}$ & $6.1 \cdot 10^{-33}$\\
			R-CM  & $2.4 \cdot 10^{-29}$ & $2.7 \cdot 10^{-33}$ & $1.7 \cdot 10^{-34}$\\
			R-K & $9.1 \cdot 10^{-27}$ & $1.1 \cdot 10^{-32}$ & - \\
			
		\end{tabular}
	\end{center}
\end{table}

We simulated the vortex decay using the \ac{SRT}, R-WO-MRT, R-CM and R-K methods (cf. \cref{subsection:Simplification}),
with all admissible combinations of storage and equilibrium format.
In \cref{fig:TGAPlot} the analytical solution of the kinetic energy is compared to simulated results.
The vortex velocity and thus the kinetic energy are expected to decay exponentially due to viscous friction.
This is matched by the simulation until, at some point, the simulated energy stagnates on a plateau.
The reason for this phenomenon is that lower velocities can no longer be represented
due to the truncation error caused by the floating point number format.
The truncation error for the IEEE-754 double precision format \cite{ieee754} is $\epsilon = 2.2 \cdot 10^{-16}$.
Due to the squared calculation for the kinetic energy, the plateau's location is expected to occur at around $\epsilon^2$.
Comparing different compute kernels derived by \lbmpy{}, we observe that some are able to reach the predicted plateau,
while others stagnate visibly earlier.
The relative energy $E\left(t\right) / E_0$ after \num{200000} time steps is also shown in \cref{tab:numericTGA}.
We consistently attain much lower levels of the plateau using zero-centered storage in combination with the
delta-equilibrium.
Relaxing zero-centered populations against the absolute equilibrium, the R-CM and R-K methods still yield very high
precision, while R-WO-MRT stagnates earlier.
For both the \ac{SRT}-, and the cumulant method, the difference between absolute and zero-centered storage is the most drastic.

\subsection{Shallow Water LBMs}
\label{subsection:ShallowWaterApplication}

The \acf{SWE} are an approximation to the \acl{NSE} for flow regimes
where the horizontal characteristic length scale is significantly larger than the vertical length scale,
making it permissible to neglect vertical flow phenomena \cite{Venturi2020CumulantSWE}.
They are obtained by integrating the \ac{NSE} in $z$-direction; fluid density is replaced by water column
height $h$ and velocity is averaged across the third dimension \cite{Zhou2002}.
The \ac{SWE} are therefore purely two-dimensional.
A number of \ac{LBM} solvers for the \ac{SWE} have been proposed \cite{Zhou2002,Rosis2017ShallowWater,Venturi2020CumulantSWE}
and succesfully applied to problems of hydraulic engineering \cite{Venturi2020FloodEvents}.
Here, we consider the central moment-based shallow water \ac{LBM} by de Rosis \cite{Rosis2017ShallowWater}
and the cumulant-based method presented by Venturi et al.\ \cite{Venturi2020CumulantSWE,Venturi2020FloodEvents}.
We show how these methods may be implemented using the \lbmpy{} modelling framework introduced in \cref{section:MRTModelling}
in just a few lines of Python code, automatically generating a highly optimized kernel implementation of the respective collision operators.
We then present simulation results obtained with the generated kernels.

The method of de Rosis is based on the discrete shallow water equilibrium proposed by Zhou \cite{Zhou2002} for the D2Q9 stencil,
which reads
\begin{equation}
    \label{eq:DiscreteShallowWaterEquilibrium}
    \begin{split}
        f_0^{\meq} &= h \left( 1 - \frac{5gh}{6} - \frac{2 \vec{u} \cdot \vec{u}}{3} \right), \\
        f_i^{\meq} &= \lambda_i h \left( 
            \frac{gh}{6}
            + \frac{\vec{\xi}_i \cdot \vec{u}}{3} 
            + \frac{(\vec{\xi}_i \cdot \vec{u})^2}{2}
            - \frac{\vec{u} \cdot \vec{u}}{3}\right) \quad (i = 1, \dots, 8).
    \end{split}
\end{equation}
Here, $g$ denotes gravitational acceleration in lattice units,
and $\lambda_i = 1$ if $\Vert \vec{\xi}_i \Vert_1 = 1$, otherwise $\lambda_i = 1/4$.
\Cref{listing:DiscreteShallowWaterEquilibrium} shows how a central moment-based method using this equilibrium
may be set up within \lbmpy{}.
In lines 3 to 13, we create an equilibrium instance from the discrete equations,
which enters the abstract representation of the shallow water method in lines 15 to 20.
We specify the same polynomial basis as used in \cite{Rosis2017ShallowWater}, fix central moments as the collision space,
and define a regularized set of relaxation rates.
Finally, line 22 invokes the code generation pipeline introduced in \cref{section:DerivationAndCodegen},
generating and compiling a C implementation of the collision kernel,
which is made available to the user as a Python function.

The method of Venturi et al.\ \cite{Venturi2020CumulantSWE} may be recreated even more compactly,
using the continuous Maxwellian and setting its squared speed of sound parameter to $c_s^2 = \frac{1}{2} g h$.
The second half of the code in \cref{listing:DiscreteShallowWaterEquilibrium} is then identical,
except for specifying \texttt{CUMULANTS} as a collision space, instead of central moments.

\begin{lstlisting}[
    float,
	floatplacement=t,
    style=pysnippet,
    firstline=11,lastline=32,
    caption={Instantiation of the discrete shallow water equilibrium using \lbmpy{}'s modelling framework;
             construction of the central moment-based shallow water \ac{LBM},
             plus automatic derivation, generation and compilation of the collision kernel.},
    label={listing:DiscreteShallowWaterEquilibrium}
]
x, y, g, h, u_x, u_y, $\omega$_s = sp.symbols(...)

def f_eq($\xi$):
    $\xi$_sum = sum(abs($\xi$_i) for $\xi$_i in $\xi$)
    if $\xi$_sum == 0:
        return h * (1 - (5 * g * h) / 6 - (2 * dot(u,u)) / 3)
    else:
        $\lambda$ = 1 if $\xi$_sum == 1 else sp.Rational(1, 4)
        common = (g * h) / 6 + dot($\xi$, u) / 3 + dot($\xi$, u)**2 / 2 - dot(u,u) / 6
        return $\lambda$ * h * common

eq_populations = [f_eq($\xi$) for $\xi$ in d2q9]
swe_eq = GenericDiscreteEquilibrium(d2q9, eq_populations, h, u)

polys = (1, x, y, x * y, x**2 - y**2, x**2 + y**2, x * y**2, y * x**2, x**2 * y**2)
r_rates = [0, 0, 0, $\omega$_s, $\omega$_s, 1, 1, 1, 1]
rr_dict = dict(zip(polys, r_rates))
cqc = DensityVelocityComputation(d2q9, density_symbol=h, ...)
cspace = CollisionSpaceInfo(CollisionSpace.CENTRAL_MOMENTS)
swe_method = create_from_equilibrium(d2q9, swe_eq, cqc, rr_dict, cspace)

lb_kernel = create_lb_function(lb_method=swe_method)
\end{lstlisting}

We put the kernels thus generated to work in simulating a circular dam break scenario, using
the same setup as in \cite{Rosis2017ShallowWater}.
We place a water column of radius $2.5 \;\mathrm{m}$ and height $2.5 \;\mathrm{m}$ at the center of a
cubic domain with side length $40 \;\mathrm{m}$, which is otherwise filled with $0.5 \;\mathrm{m}$
of water on top of an even and frictionless bed.
The domain is discretized with $100^2$ lattice cells and delimited by periodic boundary conditions;
the simulated time step is $\Delta t = 0.05 \;\mathrm{s}$ per step.
The kinematic viscosity is set to unity; this yields a shear viscosity-governing relaxation rate of
$\omega_s \approx 0.696$.
The entire simulation is set up very rapidly in Python code using the additional facilities
of \pystencils{} and \lbmpy{}.
The full Python code for this setup is available as part of \lbmpy{}'s online documentation%
\footnote{%
	\href{https://pycodegen.pages.i10git.cs.fau.de/lbmpy/notebooks/demo_shallow_water_lbm.html}{%
	pycodegen.pages.i10git.cs.fau.de/lbmpy}%
}.

\begin{figure}
    \begin{center}
        \includegraphics{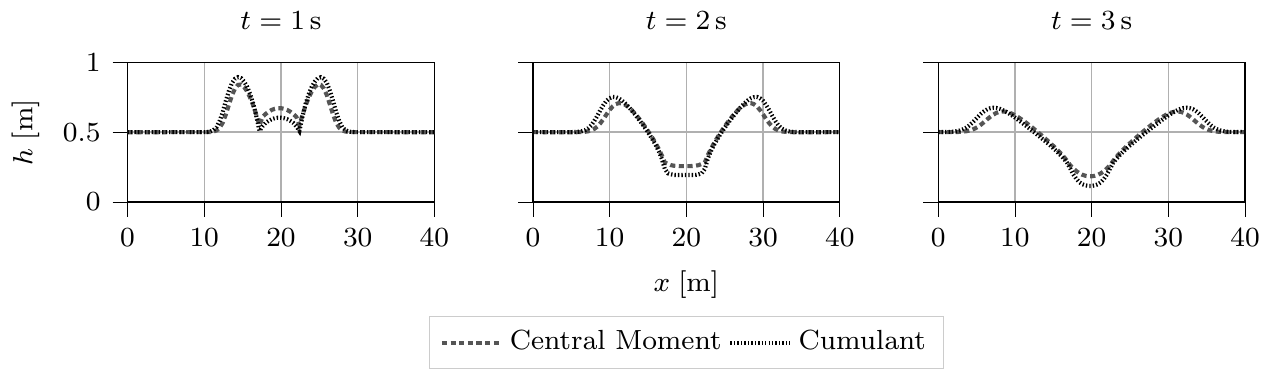}
    \end{center}
    \caption{Comparison of water depth in circular dam break simulations using the central-moment based (CM)
             and cumulant-based (K) shallow water \acp{LBM}.}
    \label{fig:CircularDamBreakPlot}
\end{figure}

\begin{figure}
	\begin{center}
		\includegraphics[width=\textwidth]{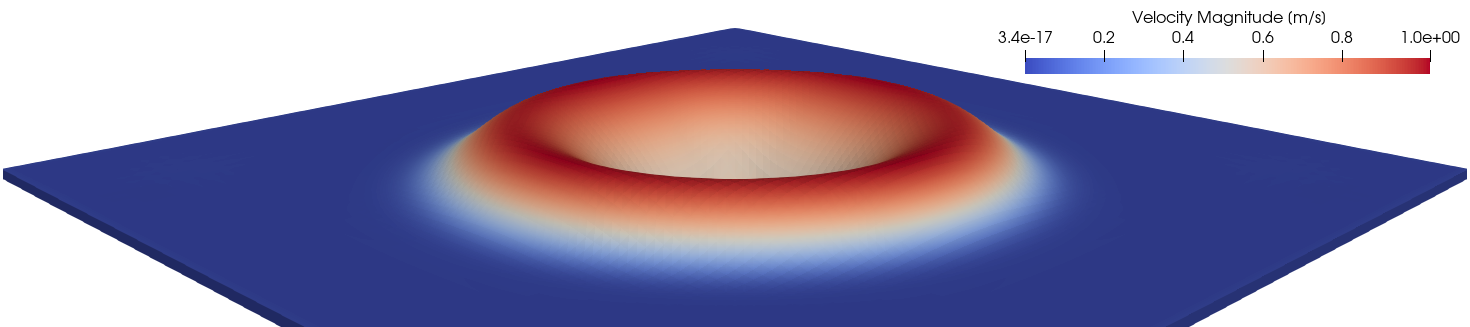}
	\end{center}
	\caption{Circular dam break simulation at $t = 2\, \mathrm{s}$ using the central moment-based method.}
	\label{fig:DamBreakVisualization}
\end{figure}

\Cref{fig:CircularDamBreakPlot} shows water column height in a cross-section of the domain at $y = 20 \;\mathrm{m}$,
at $1$, $2$ and $3$ seconds of simulated time.
While both methods agree qualitatively, the cumulant-based method shows a visibly deeper trough and
a more spread-out wave front.
The wave front after two seconds, as predicted by the central moment-based method, is visualized in \cref{fig:DamBreakVisualization}.

    \section{Conclusion and Outlook}
\label{section:Conclusion}

This article presents \lbmpy{} as a sophisticated software architecture that supports the algebraic modelling
of advanced \ac{MRT} \aclp{LBM}.
Its key feature is the ability to generate efficient computer code automatically from abstract specifications of the methods.
The methodology is based on an extensive theoretical framework for the concise description of \ac{MRT} collision rules. 
The framework formalizes the zero-centered storage format as a decomposition of the populations
into background and deviation components.
This decomposition we generalize to raw and central moment space, applying it also to the
equilibrium distribution.
Thus we arrive at three general collision equations, relaxing absolute populations against
the absolute equilibrium and zero-centered populations against both the absolute and delta-
equilibrium.

We present the front-end of \lbmpy{} as a Python-based software
incarnation of this theoretical calculus.
Utilizing a computer algebra system, the software supports manipulating \acp{LBM} on
the purely symbolic level.
We further describe how object-oriented programming can be used to represent complex
components, like the equilibrium distribution and force models.
This approach makes \lbmpy{} flexible and 
extensible.

We discuss in detail the way optimized collision rules are derived within \lbmpy{}
through symbolic manipulation.
Our derivation system generates efficient implementations for collision rules in raw moment,
central moment, and cumulant space, including all permissible combinations of storage and
equilibrium formats.
We provide specialized derivations for the collision space transforms since they constitute the most expensive
computations.
The linear mappings between populations, raw, and central moments are realized using two novel Chimera transforms:
One recursively decomposing the calculation of discrete raw moments,
and one a split-up version of a threefold binomial expansion.
Both are designed to leverage the respective transforms' recursive nature
in order to minimize the amount of arithmetic operations.
We furthermore introduce a method of deriving the transforms between central moments and cumulants
by symbolic differentiation of the respective generating functions.

Combining domain-specific knowledge with information contained in the symbolic equations,
we succeed in significantly reducing arithmetic operation counts of complex \ac{LBM} kernels.
We observe that especially the common step of regularization 
permits aggressive optimizations leading to drastically reduced computational cost.
This serves to produce remarkably low operation counts; in fact, we were able to produce
cumulant-based kernels with only between thirteen (on D3Q19) and forty percent (on D3Q27)
more operations than the respective \ac{SRT} kernel.

We conduct a test case involving decaying Taylor-Green vortex flow, showing the effectiveness of the
generalized zero-centered storage format in improving numerical accuracy.
Finally, we illustrate \lbmpy{}'s versatility and fitness for rapid prototyping by setting up an \ac{LB}
solver for the \acl{SWE} in just a few lines of Python code.

The reduction of arithmetic complexity in implementations of advanced \acp{LBM} is just one first step
toward efficient large-scale fluid dynamics simulations.
In the past, \lbmpy{}-generated kernels have already been shown to be performant, both individually on single CPUs,
as well as powering massively parallel simulations on clusters of CPUs and \acp{GPU} \cite{Bauer2021lbmpy,Holzer2021}.
With the present revision of the framework, we provide a central ingredient for more time- and energy-efficient implementations
of sophisticated lattice Boltzmann solvers employing the promising central moment and cumulant methods.
Therein, we aim to minimize time and energy consumed by both software developers and computing hardware.
Therefore, the evaluation of the performance characteristics of the generated kernels on diverse modern hardware architectures
as well as the assertion of their fitness for latest peta- and exascale supercomputers
shall be subjects of future work.

    \section*{Acknowledgements}
The authors acknowledge funding by the SCALABLE project. SCALABLE has received funding from the European Union's Horizon 2020 research and innovation programme under grant agreement No 956000.
The authors are grateful to the Deutsche Forschungsgemeinschaft (DFG, German Research Foundation) for funding projects 408062554 and 433735254.
The authors also gratefully acknowledge financial support by the Bavarian State Ministry of Science and the Arts through the Competence 
Network for Scientific High Performance Computing in Bavaria (KONWIHR).

    \appendix

    \bibliography{sources.bib}
    \bibliographystyle{siamplain}

\end{document}